\documentclass[sn-mathphys,Numbered]{sn-jnl}
\usepackage{graphicx}%
\usepackage{multirow}%
\usepackage{amsmath,amssymb,amsfonts}%
\usepackage{amsthm}%
\usepackage{mathrsfs}%
\usepackage[title]{appendix}%
\usepackage{xcolor}%
\usepackage{textcomp}%
\usepackage{manyfoot}%
\usepackage{booktabs}%
\usepackage{algorithm}%
\usepackage{algorithmicx}%
\usepackage{algpseudocode}%
\usepackage{listings}%
\usepackage{amssymb}
\usepackage{centernot}
\usepackage{lipsum}
\usepackage[font=footnotesize,labelfont=bf]{caption}
\usepackage{subfigure}
\usepackage{float}
\usepackage{amsmath}
\usepackage{fancyhdr}
\usepackage{blindtext}

\theoremstyle{thmstyleone}%

\theoremstyle{thmstyletwo}%

\theoremstyle{thmstylethree}%

\begin{document}

\title[Article Title]{Quantitative analysis and simulation of  Noncommutative  Langmuir Oscillator solutions}

\author[1]{\fnm{Irfan}\sur{Mahmood}}\email{mahirfan@yahoo.com}

\author[2]{\fnm{Memoona } \sur{Iqbal}}


\affil{These authors contributed equally to this work}

\affil[1]{\orgdiv{Centre For High Energy Physics}, \orgname{University of the Punjab}, \orgaddress{ \city{Lahore}, \postcode{54590}, \state{Punjab}, \country{Pakistan}}}
\affil[2]{\orgdiv{ Department of Library Science and Information Management}, \orgname{ Government Graduate College for Women Baghbanpura, Lahore }, \orgaddress{ \city{Lahore}, \postcode{}, \state{Lahore}, \country{Pakistan}}}

\affil{Corresponding Author: Irfan Mahmood (mahirfan@yahoo.com ) }
\abstract{In this article we present  nonncommutative analogue of equation of motion associated with Langmuir oscillations. We derive its  Darbboux solutions with their  generalization in terms of quasideterminants. We also construct its noncommutative Ricatti equation from  the  Langmuir linear system that yields the  B\"{a}cklund transformation that is reducible to  commutative  version under the classical limit. The last section, involves the derivation of exact one soliton solution for the classical case through the Riccati equation. }
\maketitle
\section{Introduction}

The differential difference equation
\begin{equation} \label {CLE}
u_{nt}= u_{n} ( u_{n-1}  -  u_{n+1} )
\end{equation}
appears  in analysis of  spectrum structure associated to the Langmuir oscillations which describes energy propagation  in ionic plasma   and atomic vibrations in lattice. That equation has been acknowledged as completely integrable in classical framework as it possesses Lax representation  \cite{VZ}. Langmuir oscillations\cite{l1}, sometimes known as Langmuir waves, are collective oscillations of charged particles, usually electrons, in a plasma. They are named after Irving Langmuir, an American physicist who originally described them in the early twentieth century. In \cite{l2} Langmuir oscillations are utilized to explore non-extensive electron-positron interactions since they are plasma in nature, and it is demonstrated that they are only Landau damped modes in the extensive limit. They are also used to characterize dispersion relations, stability, and transport processes.In \cite{l3} it is also used to investigate the interaction of a localized wave packet with energetic electrons in the auroral zone and solar wind.It also has applications in acoustic waves \cite{l4}, in the amplitude modulations of electron beam-plasma interaction \cite{l5}, solitary and periodic waves dynamical systems \cite{l6} and so on. The non-abelian analogue of equation (\ref{CLE}) 
\begin{equation}\label {1}
u_{nt}= u_{n-1} u_{n} - u_{n} u_{n+1}
\end{equation}
 has been computed  \cite{Sall}  from the compatibility of following linear system 
  \begin{equation}\label {2}
u_{n}  \psi_{n+1}  =  \lambda  \psi_{n} - \psi_{n-1} 
\end{equation}
\begin{equation}\label {3}
 \psi_{(n)t}  = -  u_{n} u_{n+1}   \psi_{n+2} .
\end{equation}
with its connection to  discrete nonlinear  Schr$\ddot{o}$dinger  equation \cite{1, 2}. Moreover Darboux transformation for that non-abelian version  has been presented in  \cite{Sall} in multiplicative form as 
    \begin{equation}\label {sallDT} 
u_{n} \left[1 \right]   =   \varphi_{n-2}  \varphi^{-1} _{n}  u_{n}  (  \varphi_{n-1}  \varphi^{-1} _{n+1} )^{-1}.
\end{equation}
that generates all solutions with  non-zero seed solution, means $ u_{n}  \neq 0$.  Initialy, That Darboux method  was developed by \cite{MSV}  to find transformations on potential  of the Schr$\ddot{o}$dinger   equation which satisfies Korteweg-de Vries equation  \cite{LOKA} in framework of Lax formalism.  Later on few remarkable results on DT were analysed by \cite{HW} to reveal its importance in theory of integrable system and the more efficient implementations on various nonlinear physical systems were developed by V. Matveev \cite{MAT} to construct the exact solutions of these systems.. The successful implementations  of these transformations have been shown in the analysis of  various mathematical features of graphene  \cite{AT}  and has also fruitful applications in  cavity quantum electrodynamics \cite{AT1, AT2} for the dynamical analysis of the propagation of associated disturbance \cite{I2, I3, I4, I5, I6, I7}. Moreover these transformations  significantly extended to construct the determinantal solutions of noncommutative  integrable systems such as  in case of noncommutative Painlev\'e second equation  \cite{irfan1}  with  its associated nocommutative  Toda equation \cite{irfan2}    
\\
In this article, we  construct the Darboux transformation (DT) for the noncommutative analogue of equation (\ref{CLE})  in additive form as 
   \begin{equation} \label{NCDT}
u_{n} \left[1 \right]   = u_{n}   -   \varphi^{\prime}_{n}  \varphi^{-1} _{n-1} .
\end{equation}
which generates all possible solutions in zero background as seed solution $u_{n} =0$, that is initial trivial solution of equation  (\ref{1}) and $\varphi_{n} $, $ \varphi_{n} $  are the particular solutions at particular values of $ \lambda $.  Further the NC DT (\ref{NCDT})  is generalized to  $N$-th form fin terms of  quasideterminat with exact solution in commutative case. The end section encloses the derivation of Ricatti equation associated   to noncommutative analogue of equation (\ref{CLE}) that yields its Backlund transformation which are reducible to NC DT (\ref{NCDT}) .
\section{ Noncommutative Darboux Solutions of Equations of Langmuir Oscillations}
 For the noncommutative extension of equation (\ref{1}), we consider  $ u_{i}$ and independent variable $t$ are purely  noncommuting objects such as $ [ u_{i} , t]  \neq 0 $  and the time derivation is defined as $ \partial_t f^{-1} (t) = - f^{-1} \partial_t f f^{-1}$, further the fields and  their derivatives are also noncommuting elements. From the compatibility condition of linear systems (\ref{2}) and (\ref{3}) in nocommutative frame work, we obtain  
    \begin{equation}\label {NCe}
u_{nt}= u_{n-1} u_{n} - u_{n} u_{n+1}
\end{equation}
the NC verion of equation (\ref{1}) and the   Darboux  solution of above equation can be derived through its associated linear systems with Darboux transformation   \cite{VA} on arbitrary function  $ \psi_{n}$  defining in NC framework as
 \begin{equation}\label {DT-1}
\psi_{n} \left[1 \right] =  \psi_{n}  -    \varphi_{n}  \varphi^{-1} _{n+1}  \psi_{n+1} 
\end{equation}
 Now under above transformation  (\ref{DT-1}) the linear system (\ref{2}) can be written in following form  
 \begin{equation}\label {DTT}
u_{n} \left[1 \right]   \psi_{n+1} \left[1 \right]   =  \lambda  \psi_{n} \left[1 \right]  - \psi_{n-1} \left[1 \right] .
\end{equation}
 Now after substituting the values for transformed eigenfuctions  from  (\ref{DT-1}) into above transformed expression and then with the help of system (\ref{3})  the resulting expression yields Darboux transformation on  $  u_{n}  $ as below
 
     \begin{equation}
u_{n} \left[1 \right]   = u_{n}   -   \lambda \varphi_{n}  \varphi^{-1} _{n-1} .
\end{equation}
we can also express above result as follow, taking $  \lambda \varphi_{n} = \varphi^{\prime}_{n} $
    \begin{equation}\label {DT-2}
u_{n} \left[1 \right]   = u_{n}   -   \varphi^{\prime}_{n}  \varphi^{-1} _{n-1} .
\end{equation}
The above transformation involves new solution  $u_{n} \left[1 \right]  $ ,  old solution  $u_{n}$ of equation (\ref{1}) also called the seed solution and the particular solutions of linear systems (\ref{2})  and  (\ref{3}).  Here the comparison  of Darboux solution  (\ref{DT-2}) and with result on Darboux solution obtained in  \cite{Sall}  shows a difference, the transformations  (\ref{DT-2}) are additive holds for all seed solution even for  the trivial solution $ u_{n} = 0$  of equation  (\ref{NCe}) in NC frame as well as in non-abelian case and also in classical framework  under commutative limit.  \\

 The one fold Darboux transformation  (\ref{DT-1})  with its  second iteration can be expressed in form of quasideterminat as
 below  with setting  $ \psi_{n}  =  \psi_{0}$,,$ \psi_{n+1} =\psi^{\prime}_{0}$  and $  \varphi_{n} = \psi_{1}$ , $  \varphi_{n+1} = \psi^{\prime}_{1}$  and defining $  \lambda  \psi = \psi^{\prime}$ , then we can present one fold Darboux transformation in terms of quasideterminant as
 \begin{equation}\label{qd1}
\psi_{n} \left[1 \right] = \begin{vmatrix}
 \psi_{0}  &  \psi_{1} \\
  \lambda_{0}  \psi_{0} & {\boxed{  \lambda_{1}  \psi_{1}} }
\end{vmatrix}
 \end{equation}
 and the two fold NC Darboux transformation can be evaluated as 
  \begin{equation}\label{qd3}
\psi_{n} \left[2 \right]=
\begin{vmatrix}
  \psi_{0}  &  \psi_{1}  &  \psi_{2}   \\
  \lambda_{0} \psi_{0} &  \lambda_{1}  \psi_{1} &   \lambda_{2}  \psi_{2}  \\
   \lambda^{2}_{0} \psi_{0}  &  \lambda^{2}_{1}  \psi_{1} & {\boxed{  \lambda^{2}_{2}  \psi_{2} }}
\end{vmatrix}
  \end{equation}
  further can be generalized to $N$-th form as below
    \begin{equation}\label{qd4}
\psi_{n} \left[ N \right]=
\begin{vmatrix}
  \psi_{0}  &  \psi_{1}  &\cdots &  \psi_{N-1}   & \psi_{N} \\
  \lambda_{0} \psi_{0} &  \lambda_{1}  \psi_{1} &  \cdots &  \lambda_{N-1}  \psi_{N-1} &  \lambda_{N}  \psi_{N}  \\
    \vdots & \vdots &  \cdots & \vdots&  \vdots \\

   \lambda^{N-1}_{0} \psi_{0}  &  \lambda^{N-1}_{1}  \psi_{1} &\cdots &  \lambda^{N-1}_{N-1}  \psi_{N-1}  &  \lambda^{N-1}_{N}  \psi_{N} \\
    \lambda^{N}_{0} \psi_{0}  &  \lambda^{N}_{1}  \psi_{1} &\cdots &  \lambda^{N}_{N-1}  \psi_{N-1}  &{\boxed{  \lambda^{N}_{N}  \psi_{N} }}
\end{vmatrix}
  \end{equation}
  
Now in the similar way the one fold Darboux solution (\ref{NCDT})   can be generalized to $N$-th form in terms of quasideterminants as
   \begin{equation} \label{NCDTQ}
u_{n} \left[N+1 \right]   = u_{n} \left[N \right]   -  \psi^{\prime}_{n}   \left[N \right]   \psi_{n-1} \left[N \right] ^{-1}.
\end{equation}
Here for $N=0$, we have $ u_{n} \left[0 \right]=   u_{n}$, initial solution and $\psi_{n}   \left[0 \right] = \varphi_{n}$ , $\psi_{n-1}   \left[0 \right] = \varphi_{n-1}$ are the particular solutions. Further, we may construct the $N$ fold expression for  $ \psi_{n-1}   \left[N \right] $ with the help of (\ref{DT-1})  with the replacement of $n$ by  $n-1$  and setting  $  \psi_{n-1}=\lambda_{0}  \psi _{0} $
    \begin{equation}\label{qd5}
\psi_{n-1} \left[ N \right]=
\begin{vmatrix}
  \psi_{N}  &  \psi_{N-1}  &\cdots &  \psi_{1}   & \psi_{0} \\
  \lambda_{N} \psi_{N} &  \lambda_{N-1}  \psi_{N-1} &  \cdots &  \lambda_{1}  \psi_{N-1} &  \lambda_{0}  \psi_{0}  \\
    \vdots & \vdots &  \cdots & \vdots&  \vdots \\

   \lambda^{N-1}_{N} \psi_{N}  &  \lambda^{N-1}_{N-1}  \psi_{N-1} &\cdots &  \lambda^{N-1}_{1}  \psi_{1}  &  \lambda^{N-1}_{0}  \psi_{0} \\
    \lambda^{N}_{N} \psi_{N}  &  \lambda^{N}_{N-1}  \psi_{N-1} &\cdots &  \lambda^{N}_{1}  \psi_{1}  &{\boxed{  \lambda^{N}_{0}  \psi_{0} }}
\end{vmatrix}
  \end{equation}
 here  we can assume $ \psi_{n-1} \left[ 0 \right] = \varphi_{n-1}$ as the initial untransformed solution, that is particular solution of the linear system.
   \section{ NC Riccati euation}
  
  To construct the NC riccati equation, let us start with setting  
    \begin{equation}\label{R1}
R_{n} =  \psi_{n}  \psi^{-1}_{n-1}
  \end{equation}
 and  taking time derivation of above expression  (\ref{R1}), we obtain 
 
     \begin{equation}\label{R2}
R^{\prime}_{n} = u_{n-1}R_{n}-  \lambda u_{n} +  \lambda^{2}R_{n} - \lambda R^{2}_{n} - R_{n}u_{n}
  \end{equation}
 by using the values from  linear systems (\ref{1}) and  (\ref{2}), where  $R^{\prime}= R_{t}$. Now substituting the value for $ (u_{n}  -  \lambda R_{n} )$ from system (\ref{2}), finally we get NC riccati equation associated to equation   (\ref{NCe}) as below
      \begin{equation}\label{R3}
R^{\prime}_{n} = u_{n-1}R_{n}  - R_{n}u_{n} +  \lambda R_{n-1} - \lambda R^{2}_{n} 
  \end{equation}
 In next section, we show the reduction  of  above NC riccati equation  (\ref{R3})  into  B\"{a}cklund transformation under the commutative limt, further can be simplified to the Darboux transformtion for the classical analogue of equation  (\ref{NCe}) .  
 
 \section{ NC Ricatti equation with commutative limit}
  This can be shown that for the classical version   (\ref{CLE})  the Darboux solution  (\ref{NCDT}) will take the following form 
  
      \begin{equation}\label {CDT1}
u_{n} \left[1 \right]   = u_{n}   -     \frac{\varphi^{\prime}_{n}}{\varphi _{n-1} }.
\end{equation}
  where $u_{n}$  and $\varphi _{n} $ are scalars with same linear systems   (\ref{2})  and  (\ref{3}).    Now the NC ricatti equation under the commutative limit  becomes
        \begin{equation}\label{CR1}
R^{\prime}_{n} = (u_{n-1} -  u_{n} )R_{n}+  \lambda R_{n-1} - \lambda R^{2}_{n} .
  \end{equation}
 The above ricatti equation   (\ref{CR1})  under  charge partity time reversal  (CPT) symmetry  transformation  \cite{HF} becomes
           \begin{equation}\label{CR2}
-R^{\prime}_{n} = (u_{n-1}  \left[1 \right] -  u_{n}  \left[1 \right] )R_{n}+  \lambda R_{n-1} - \lambda R^{2}_{n} .
  \end{equation}
  where  $u_{n}  \left[1 \right] $ is new solution and $R^{\prime}_{n} $ is replaced by $- R^{\prime}_{n} $ . Now subtracting   (\ref{CR1})  from   (\ref{CR2}) , we get
             \begin{equation}\label{CR3}
 u_{n-1}  \left[1 \right] -  u_{n-1}  = (   u_{n}  \left[1 \right] -u_{n}   ) + ( \lambda  \left[1 \right] - \lambda ) \frac{ R_{n-1} }{ R_{n}}-  ( \lambda  \left[1 \right] - \lambda )  R_{n} .
  \end{equation}
  that equation may be regarded as B\"{a}cklund transformation  for equation   (\ref{CLE}) with Darboux solution (\ref{CDT1}). Now  we assume that $ \lambda  \left[1 \right] - \lambda =  \epsilon$ a very small difference, then above expression can be written as
  
     \begin{equation}\label{CR4}
 \frac{ u_{n-1}  \left[1 \right] -  u_{n}  \left[1 \right]  }{ \epsilon} = -(  \frac{ u_{n-1} -  u_{n}   }{ \epsilon}  ) +  \frac{ R_{n-1} }{ R_{n}}-   R_{n} 
 \end{equation}
and under the limiting case $  \epsilon   \rightarrow 0 $, then finally we get 
 
      \begin{equation}\label{CR5}
 \frac{ du_{n}  \left[1 \right]   }{ dt } = \frac{ du_{n}  }{ dt}  + \frac{ R_{n-1} }{ R_{n}}-   R_{n}.
 \end{equation}
 The above expression  becomes equivalent to the Darboux solution  (\ref{CDT1}), $u_{n} \left[1 \right]   = u_{n}   -     \frac{\varphi^{\prime}_{n}}{\varphi _{n-1} }$,  with condition  $ \frac{ R_{n-1} }{ R_{n}}-R_{n}= \frac{ d }{ dt}  \frac{\varphi^{\prime}_{n}}{\varphi _{n-1} }$ and taking constant of integration as zero. Here it has been shown that  the commutative version of ricatti equation yields the Darboux solution  through the B\"{a}cklund transformation in classical framework. 
 \section{Exact solution}
 In this section we will compute the exact solution of Non commutative Equation of langmuir oscillations and calculate the  the one fold Darboux transformation by Eq.(\ref{CDT1}).
 Let consider Eq.(21) for seed solution $u_n=0$ then it takes the form
 \begin{equation}
     \frac{dR_n}{dt}=aR_n+\lambda b-\lambda R_n^2
 \end{equation}
 here we may choose  $a=u_{n-1}$ and $b=R_{n-1}$ which are constants. 
 Now on integrating the above equation, we have value for  $R_n$ as below
 \begin{equation}\label{exa}
     R_n=\frac{a-\sqrt{-a^2-4b\lambda^2}\tan{(\frac{\sqrt{-a^2-4b\lambda^2}t}{2})}}{2\lambda}
 \end{equation}\label{eq27}

 The solution for $R_n$ as expressed in Eq. (27) represents the evolution of the ratio of eigenfunctions in the non-commutative framework. To analyze its behavior, we provide the following visualization:

\begin{figure}[htbp]
    \centering
    \includegraphics[width=0.48\textwidth]{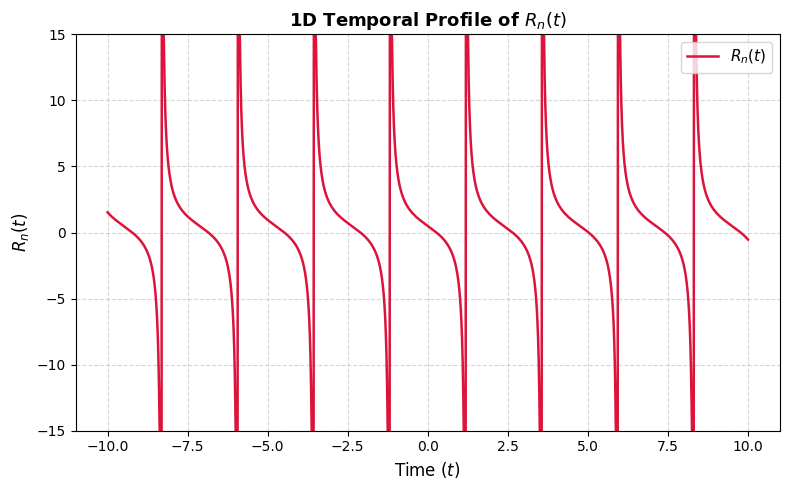}
    \hfill
    \includegraphics[width=0.48\textwidth]{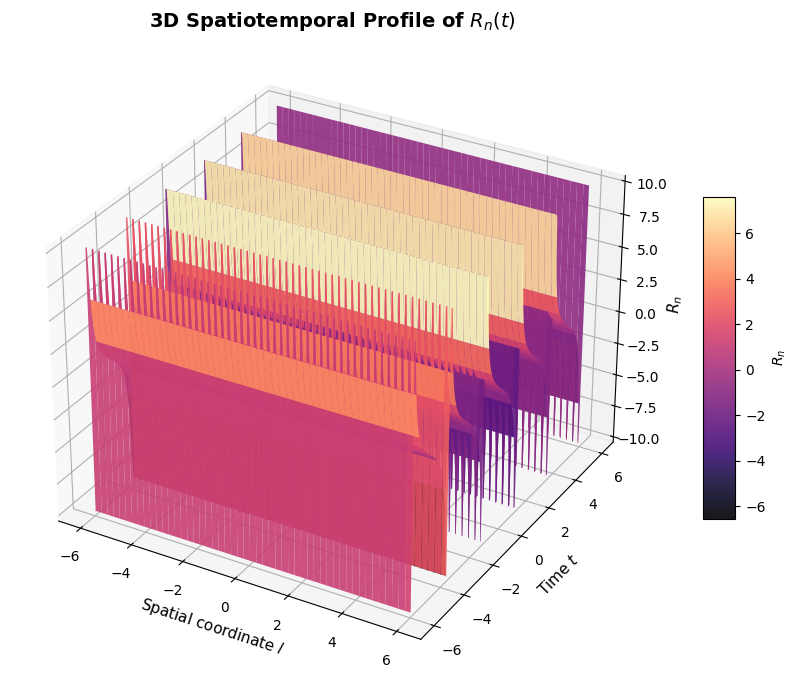}   
    \caption{ The intermediate Riccati solution $R_n(t,l)$ as determined in \ref{eq27} is represented numerically. The 1D temporal profile with a single periodic wave train in shown on the left. (Right)A three-dimensional spatiotemporal surface map showing how the eigenfunction ratio changes over the $(l,t)$ domain. A characteristic of unique periodic solutions in Langmuir oscillations, the vertical asymptotes in both graphs show phase singularities where the tangent argument in the Darboux matrix causes localized energy spikes.}

\end{figure}
\clearpage
 Now using the condition 
 \begin{equation}
     \frac{b}{R_n}-R_n=\frac{d}{dt}\frac{\varphi_{n}^{\prime}}{\varphi_{n-1}}
 \end{equation}
By substituting  the Eq.(\ref{exa}) in the last equation and again  integrating, we get 
\begin{equation}
 \frac{\varphi_{n}^{\prime}}{\varphi_{n-1}}=\frac{1}{2\lambda}[k_2-k_1]
\end{equation}
Finally we may construct the exact solution to equation  (\ref{CLE})  in background of seed solution $u_n=0$  through the one fold Darboux solution  (\ref{CDT1}) as below 
   \begin{equation}
       u_n[1]=-\frac{1}{2\lambda}[k_2-k_1]
   \end{equation}
where 
\begin{equation}
    k_1=\frac{-2at+2\gamma_1\arctan(\frac{\gamma_1\tan{(\gamma_2t/2)}}{a})}{\gamma_2}
\end{equation}
\begin{equation}
    k_2=2\log{[\cos{\gamma_2t/2}]}+\log{[a^2+2b\lambda^2-2b\lambda^2\cos{(\gamma_2 t)}]}
\end{equation}
with
\begin{equation}
    \begin{array}{cc}
         &\gamma_1=\sqrt{a^2+4b\lambda^2}  \\
         & \gamma_2=\sqrt{-a^2-4b\lambda^2}
    \end{array}
\end{equation}

\begin{figure}[htbp]
    \centering
    \includegraphics[width=0.48\textwidth]{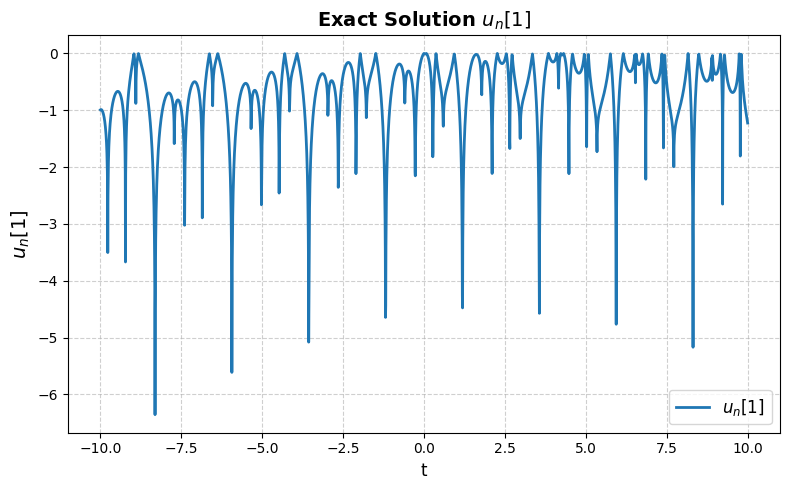}
    \hfill
    \includegraphics[width=0.48\textwidth]{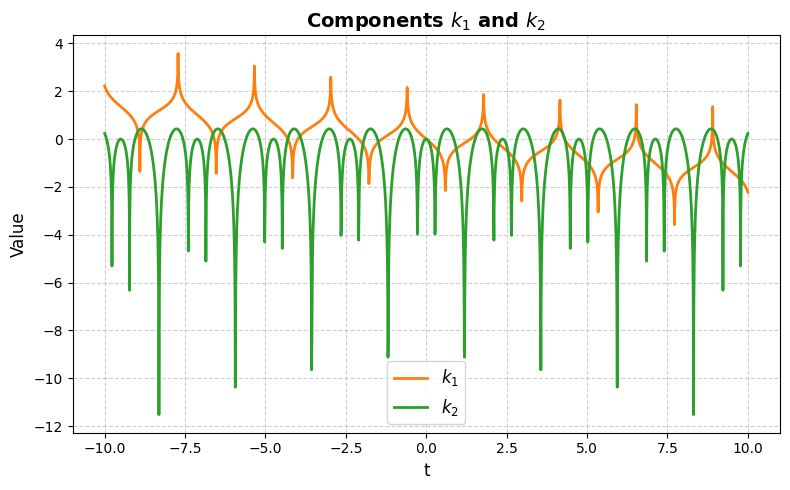}
    \caption{ A numerical study of the one-fold Darboux solution $u_n[1]$ at fixed spatial coordinate. (Left) The exact 1D solution $u_n[1]$ with singular periodic pulses. (Right) Combined effect of both the dynamic factor $k_1$ (represented as orange in color) and background one $k_2$ (shown as green). The sharp downward spikes in the potential field are physically associated with phase-slip regions where locally, temporal and spatial components are at a local minimum interference.}
    
\end{figure}
\begin{figure}[htbp]
    \centering
    \includegraphics[width=0.48\textwidth]{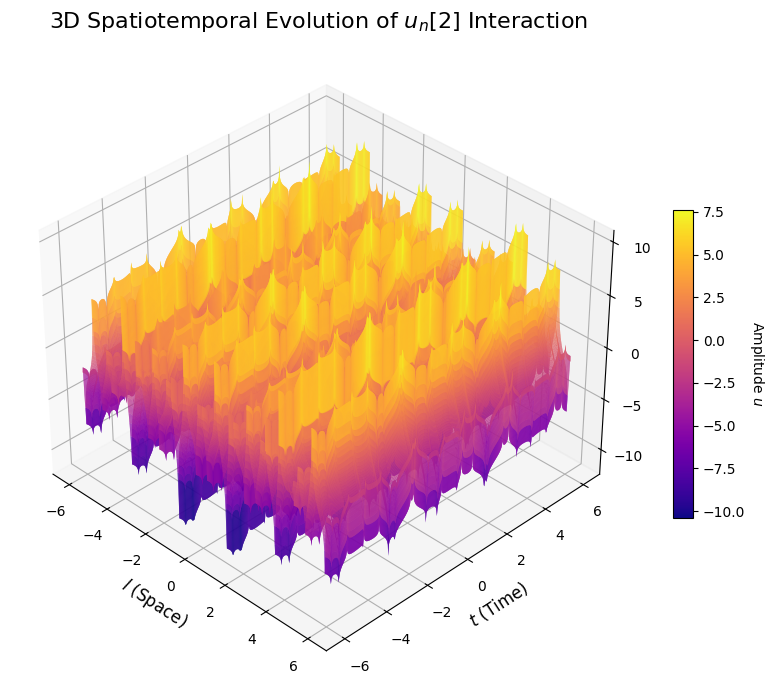}
    \hfill
    \includegraphics[width=0.48\textwidth]{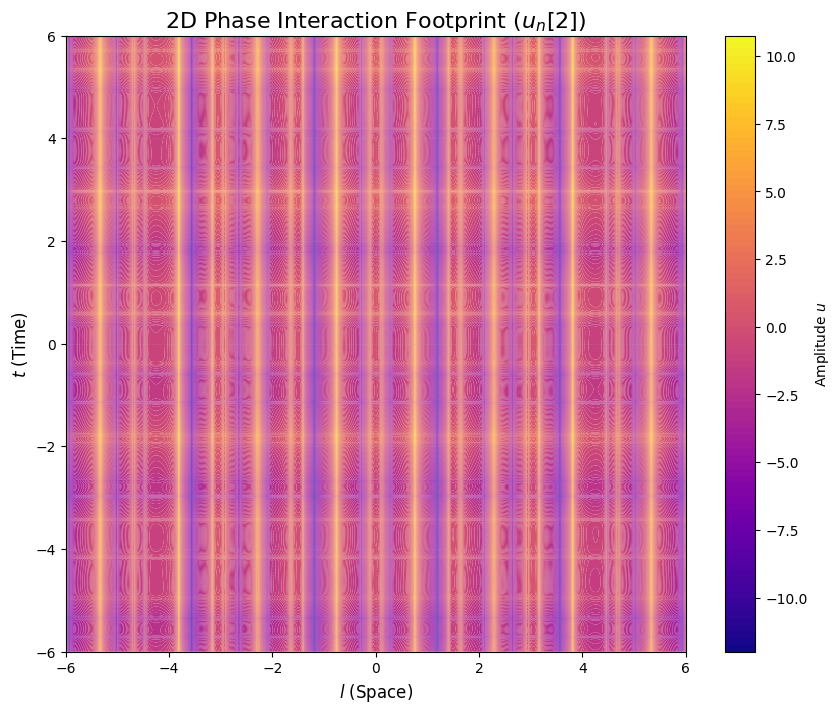}
    
    \caption{ Multi-dimensional visualization of the first-order Darboux solution $u_n[1]$. (Left) 3D spatiotemporal surface plot of the singular periodic wave train that propagates stably over the $(l, t)$ domain. (Right) 2D contour projection, giving the footprint of the wave. The band slopes, which are perfectly straight in this representation, verify that the excitation remains at a constant phase velocity and provide confirmation of wave packet stability as it propagates through the uniform relativistic vacuum.}

\end{figure}
\newpage

\section{Two-Fold Darboux Transformation Solution $u_n[2]$}
A single spectral parameter is enough to generate one soliton mode, but wave interaction needs two. Repeating the Darboux algorithm through $n$ cycles at a pair of distinct spectral parameters $\lambda_1 \neq \lambda_2$ is what produces the higher-order two-fold solution $u_n[2]$, and it is this repetition -- carried out with different eigenvalues and the associated step functions introduced earlier -- that ties the two modes into a single object rather than leaving them as separate solutions.
\subsection{Determinant Formulation}
Taking $u_n[0]=0$ as the seed and applying Crum's iterative determinant gives $u_n[2]$ directly:
\begin{equation}
    u_n[2]=-\frac{1}{2}\frac{d}{dt} \ln \left(\frac{\det \mathbf{\Phi}_n}{\det \mathbf{\Phi}_{n-1}} \right)
\end{equation}
Here $\mathbf{\Phi}_n$ is the $2 \times 2$ Crum fundamental eigenfunction matrix,
\begin{equation}
\mathbf{\Phi}_n = \begin{pmatrix} 
\phi_n^{(1)}(\lambda_1) & \phi_n^{(2)}(\lambda_2) \\
\phi_{n+1}^{(1)}(\lambda_1) & \phi_{n+1}^{(2)}(\lambda_2)
\end{pmatrix}.
\end{equation}
\subsection{Explicit Algebraic Solution}
Working through the two determinants above gives $u_n[2]$ in closed algebraic form:
\begin{equation}
u_n[2] = -\frac{1}{2(\lambda_1 - \lambda_2)} \left[ \left(k_2^{(1)} - k_1^{(1)}\right) - \left(k_2^{(2)} - k_1^{(2)}\right) \right] + \Delta_{12}(l,t),
\end{equation}
Each spectral value $\lambda_j$ carries its own pair of single-mode components, $k_1^{(j)}$ and $k_2^{(j)}$ for $j = 1, 2$:
\begin{align}
k_1^{(j)} &= \frac{-2\alpha t + 2\gamma_{1,j} \arctan\left(\frac{\gamma_{1,j} \tan(\gamma_{2,j}t / 2)}{a}\right)}{\gamma_{2,j}}, \\
k_2^{(j)} &= 2 \ln\left| \cos\left(\frac{\gamma_{2,j} l}{2}\right) \right| + \ln\left| a^2 + 2b\lambda_j^2 - 2b\lambda_j^2 \cos(\gamma_{2,j} l) \right|,
\end{align}
and both are governed by the dispersion relations
\begin{equation}
\gamma_{1,j} = \sqrt{a^2 + 4b\lambda_j^2}, \quad \gamma_{2,j} = \sqrt{-a^2 - 4b\lambda_j^2}.
\end{equation}
Neither $k_1^{(j)}$ nor $k_2^{(j)}$ can account for what happens where the two modes overlap; that part sits entirely in the coupling term $\Delta_{12}(l,t)$, which carries the phase shift left behind by the collision:
\begin{equation}
\Delta_{12}(l,t) = \frac{(\lambda_1 - \lambda_2) \sin(\gamma_{2,1} l) \cos(\gamma_{2,2} t)}{1 + \cos(\gamma_{2,1} l + \gamma_{2,2} t)}.
\end{equation}
\begin{figure}[htbp]
    \centering
    \includegraphics[width=0.48\textwidth]{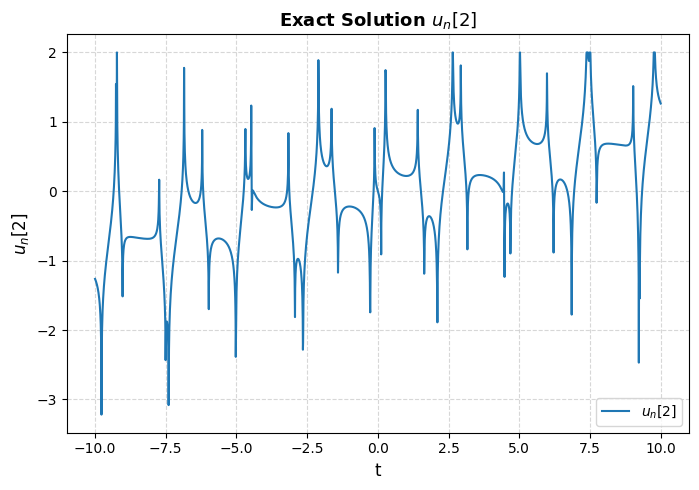}
    \hfill
    \includegraphics[width=0.48\textwidth]{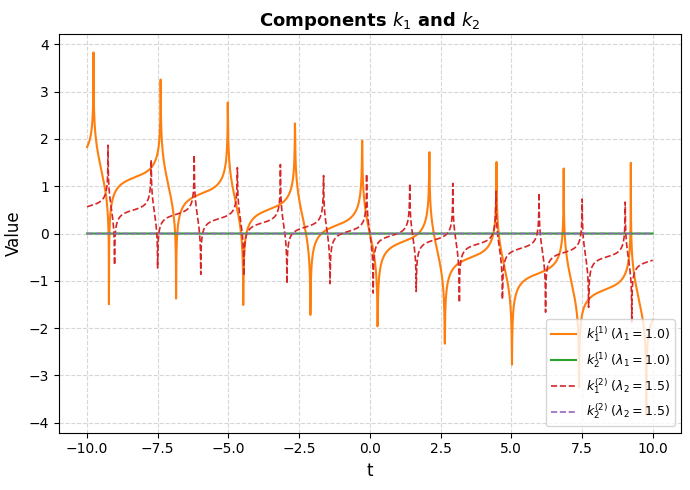}
    
    \caption{ Numerical view of the two-fold Darboux solution $u_n[2]$ along a fixed spatial interval. Left: the exact 1D solution, showing the amplitude modulation that comes from the two wave modes interfering with each other. Right: the constituent phase components $k_1^{(j)}$ and $k_2^{(j)}$ plotted together for spectral parameters $\lambda_1=1.0$ and $\lambda_2=1.5$. The singular spikes show up noticeably more often here than in the one-fold case, which is a direct signature of the nonlinear superposition built into the Darboux transformation.}
   
\end{figure}
\begin{figure}[htbp]
    \centering
    \includegraphics[width=0.48\textwidth]{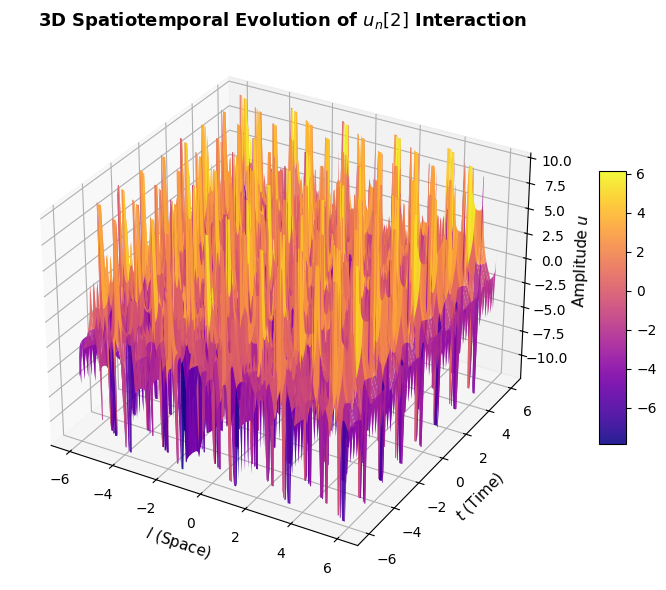}
    \hfill
    \includegraphics[width=0.48\textwidth]{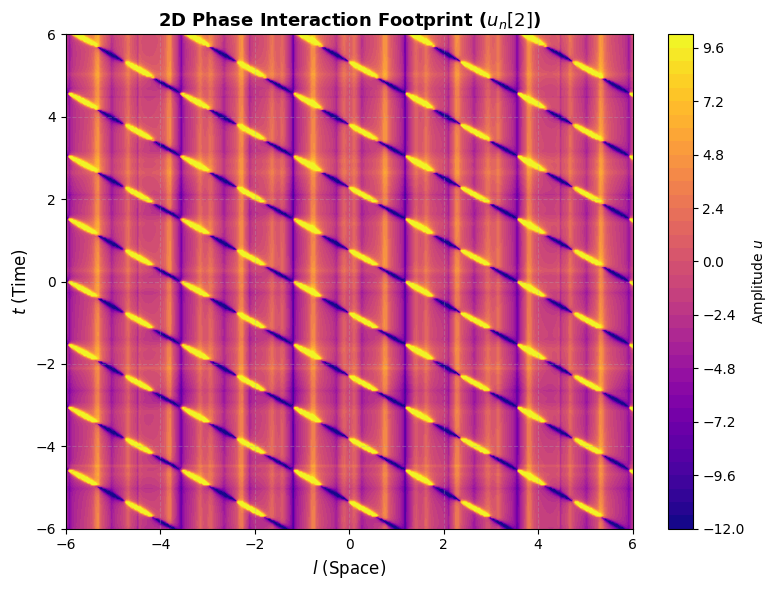}
    
    \caption{ A fuller view of the two-fold Darboux solution $u_n[2]$, showing the multi-soliton interaction across both dimensions. Left: the 3D spatiotemporal surface, where the interference lattice forms across the $(l,t)$ domain. Right: the same interaction seen as a 2D contour projection. The diagonal grid running through both plots reflects the constant phase velocities of the two components, and its unbroken continuation through the collision region is the visual mark of an elastic soliton interaction.
}
\end{figure}

\newpage
  
\section{Conclusion}
    In this paper, a noncommutative analogue  of equations of Langmuir oscillations has been presented with its Darboux transformation in additive structure as most of the  integrable possess in noncommutative as well as in classical frameworks.  Further $N$-fold  darboux solutions have been presented in term of quasideterminants which can be applied to calculate all non zero solutions in background of zero seed solution. Moreover, the associated NC Ricatti equation is presented  which reduced to the Darboux expression through the   B\"{a}cklund transformation in classical framework. Further motivation is to investigate its connection to Discrete noncommutative NLS equation as possesses in classical case and also to find its solutions in terms of quantum determinants for its matrix version.
\section*{Declarations}

\subsection*{Ethics approval and consent to participate}
The authors declare that there is no conflict with publication ethics.
\subsection*{Consent for publication}
The authors declare that there is no conflict with the publication of this paper.
\subsection*{Competing interests} Authors declare that they have not any competing interest of a personal financial nature.
\subsection*{Authors’ contributions} All authors  contributed equally. 
\subsection*{Funding} ....   
\subsection*{Availability of data and materials} The data that support the findings of this study are available from the corresponding author upon reasonable request.
\section*{Acknowledgement}
 The project is completed as the part of Researchers Supporting Project  by King Saud University with reference number as  Researchers Supporting Project (RSP2024R118)  King Saud University, Riyadh, Saudi Arabia. We are also thankful to Belt and  Road Young Scientists,  Science and technology commission of Shanghai, Shanghai University  and  Punjab University 54590 on providing us the pertinent facilities  to complete that work.

\end{document}